# Layer-Polarized Anomalous Hall Effect in Valleytronic van der Waals Bilayers

Ting Zhang,[1] Xilong Xu,[1] Baibiao Huang,[1] Ying Dai,[1*] Liangzhi Kou,[2] Yandong Ma[1*]

[1]School of Physics, State Key Laboratory of Crystal Materials, Shandong University, Shandanan Str. 27, Jinan 250100, People's Republic of China

[2]School of Mechanical, Medical and Process Engineering, Queensland University of Technology, Garden Point Campus, Brisbane, Queensland 4001, Australia

Layer-polarized anomalous Hall effect (LP-AHE), derived from the coupling between Berry curvature and layer degree of freedom, is of importance for both fundamental physics and device applications. Nonetheless, the current research paradigm is rooted in topological systems, rendering such phenomenon rather scarce. Here, through model analysis, we propose an alternative, but general mechanism to realize the LP-AHE in valleytronic van der Waals bilayers by interlayer sliding. The interaction between the out-of-plane ferroelectricity and A-type antiferromagnetism gives rise to the layer-locked Berry curvature and thus the long-sought LP-AHE in the bilayer systems. The LP-AHE can be strongly coupled with sliding ferroelectricity, to enable ferroelectrically controllable and reversible. The mechanism is demonstrated in a series of real valleytronic materials, including bilayer $VSi_2P_4$, $VSi_2N_4$, $FeCl_2$, $RuBr_2$ and $VClBr$. The new mechanism and phenomena provide a significant new direction to realize LP-AHE and explore its application in electronics.

**Introduction**

Berry curvature is a fundamental physical quantity in the field of condensed-matter physics quantifying the local entanglement of Bloch electrons under either time-reversal (*T*) or inversion (*P*) symmetry breaking [1]. In crystalline solids, it can be integrated with other quantities and thereby governs the electronic properties. This leads to a great deal of unconventional phenomena and possibilities for physical applications [2-7]. Typical examples include the quantum anomalous Hall effect (QAHE) [3,8-11] and valley Hall effect (VHE) [4,7,12-16]. The former combines Berry curvature with *T*-symmetry broken related spin degree, enabling the charge transporting via a dissipationless chiral-edge channel [17,18]. The latter links Berry curvature with *P*-symmetry broken related valley degree, which gives rise to the intriguing valley physics and valleytronics [13,19]. To date, these Berry curvature related physics has ignited tremendous research interests [1-19].

Other than the electronic degree of freedom in momentum space, real space degree that correlates to Berry curvature also plays an important role in today's condensed-matter physics. For example, a novel striking layer Hall effect (LHE), wherein the electrons from the top and bottom layers spontaneously deflect in opposite directions, could arise with Berry curvature coupling to the layer degree of freedom in van der Waals multilayers [20]. When an electric field is applied, the layer-polarized anomalous Hall effect (LP-AHE) can emerge [20]. Similar to QAHE and VHE, LP-AHE is particularly promising for novel physics and device applications. Though highly valuable, the current research on LP-AHE is exclusively rooted in the paradigm of topological physics [20]. This makes the LP-AHE rather scarce up to date. Evidently, to conquer this challenge, one must go beyond the existing paradigm to realize the LP-AHE.

In this work, we propose a different mechanism to realize the LP-AHE based on the paradigm of valleytronic van der Waals bilayers. Using model analysis, we reveal that the sliding ferroelectricity will couple with A-type antiferromagnetism to form the layer-locked Berry curvature in the bilayer systems, endowing the observation of the long-sought LP-AHE. The intriguing phenomenon can be demonstrated from the ferroelectric control of LP-AVH. The symmetry requirements are mapped out. More importantly, based on first-principles calculations, we further demonstrate the validity of this new mechanism in a series of real valleytronic materials, including bilayer $VSi_2P_4$, $VSi_2N_4$, $FeCl_2$, $RuBr_2$ and VClBr. These insights and phenomena greatly enrich the research on LP-AHE.

Our first-principles calculations are performed based on the density functional theory (DFT) [21] using the projector augmented wave method as implemented in the Vienna ab initio simulation package (VASP) [22]. The generalized gradient approximation (GGA) in the form of Perdew-Burke-Ernzerhof (PBE) [23] is adopted for the exchange correlation. The energy and force convergence criteria are set to $10^{-5}$ eV and 0.01 eV/Å, respectively. The vacuum space is set to more than 20 Å to avoid adjacent interactions. The cutoff energy is set to 550 eV. To overcome the problems of band-gap underestimation in PBE functional, we adopt the Heyd-Scuseria-Ernzerhof (HSE06) hybrid functional [24] for band structure calculations. The Brillouin zone is sampled using a $15 \times 15 \times 1$ Monkhorst-Pack k-point mesh for PBE calculations, and $7 \times 7 \times 1$ for HSE06 calculations. The Grimme's DFT-D3 method is employed [25]. We employ VASPBERRY to calculate the Berry curvature [26]. The ferroelectric polarization is evaluated using the Berry phase approach [27]. The energy barrier of ferroelectric switching is investigated using the nudged elastic band (NEB) method [28].

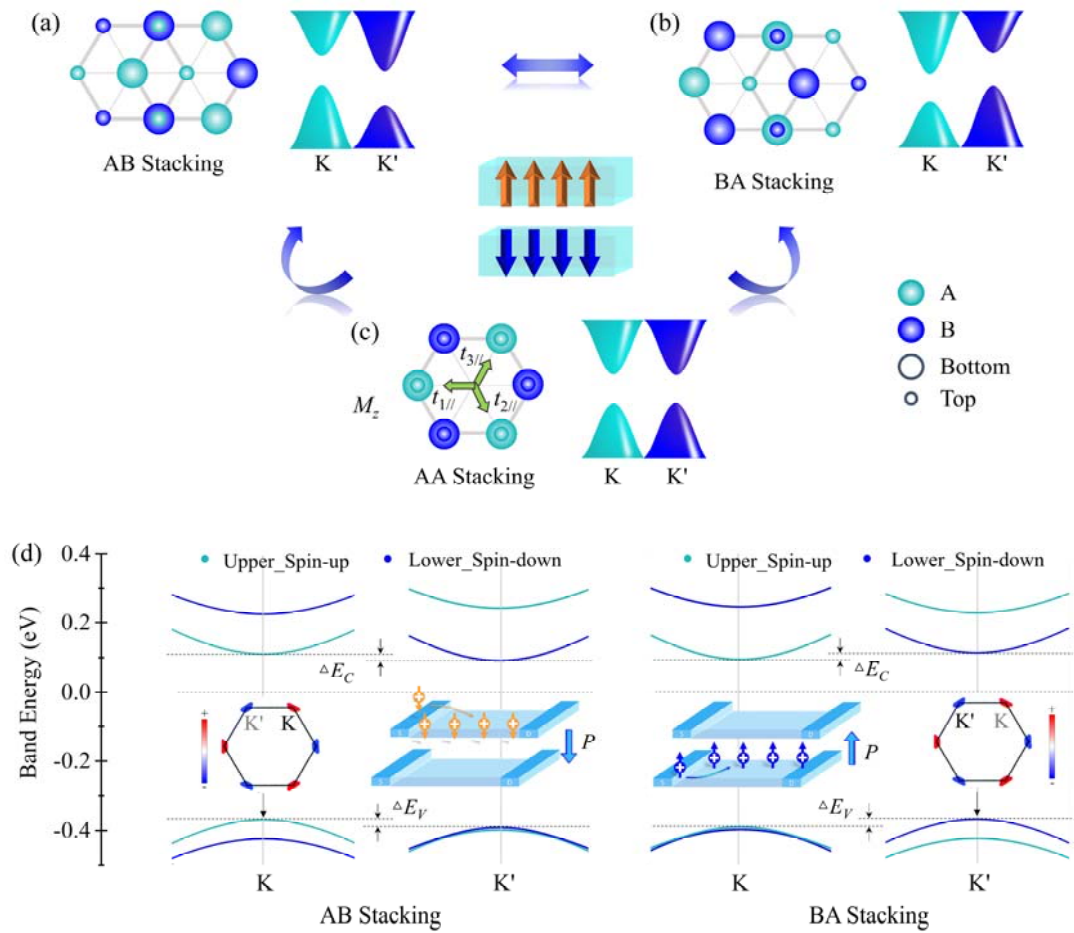

**Fig. 1** Schematic diagrams of (a) AB, (b) BA and (c) AA patterns of a bilayer lattice and their low-energy band dispersions around the K and K' valleys. Light blue and dark blue cones in (a-c) correspond to spin-up bands from upper layer and spin-down bands from lower layer, respectively.

(d) Low-energy band dispersions around the K and K' valleys for AB and BA patterns obtained from *k·p* model. Schematic representations of the Berry curvature and LP-AHE are plotted in the insets of (d).

The proposed mechanism starts from single-layer systems with spontaneous valley polarization. Such systems normally exhibit ferromagnetic (FM) exchange interaction and a hexagonal lattice with space group $P\bar{6}m2$ [29-33]. Therefore, both *T*-symmetry and *P*-symmetry are broken. We then stack two single layers together in AA pattern to construct a bilayer lattice. In the bilayer lattice, antiferromagnetism normally dominates the interlayer exchange interaction because a competition between the different interlayer orbital hybridizations [34]. This, combined with the protection of $M_z$ symmetry, prohibits the spontaneous valley polarization, as shown in **Fig. 1(c)**. For the resultant bilayer lattice, large Berry curvatures with opposite signs would be obtained around the K and K' valleys respectively. Since the K and K' valleys are from the spin-up channel of upper layer and spin-down channel of lower layer, respectively, the layer-locked spin and Berry curvature could be realized in the bilayer lattice; see **Fig. 1(c)**.

Under an interlayer translation operation of $t_{1//}\left[-\frac{2}{3},-\frac{1}{3},0\right]$, $t_{2//}\left[\frac{1}{3},-\frac{1}{3},0\right]$ or $t_{3//}\left[\frac{1}{3},\frac{2}{3},0\right]$ (-$t_{1//}$, -$t_{2//}$ or -$t_{3//}$), the AA pattern could be transformed into AB (BA) pattern, when $M_z$ symmetry is broken (space group *P*3*m*1), as shown in **Figs. 1(a)** and **(b)**. In AB and BA patterns, the absence of $M_z$ symmetry would give rise to an out-of-plane electric polarization. As illustrated in **Figs. 1(a)** and **(b)**, the existence of electric polarization can break the energetical degeneracy between the K and K' valleys. It is thus possible to manipulate the Berry curvature only at a given layer, to facilitate the realization of LP-AVH. Moreover, AB and BA patterns are energetically degenerate with opposite electric polarizations, which can be regarded as two ferroelectric (FE) states. These two FE states can be reversibly switched under interlayer sliding, leading to the intriguing sliding ferroelectricity. Upon the ferroelectric switching, as shown in **Figs. 1(a)** and **(b)**, the electric polarization and the sign of valley polarization can be simultaneously reversed, holding the potential for FE control of layer-locked physics.

To confirm this hypothesis, we employ an effective *k·p* model to characterize the low-energy band dispersions around the Fermi level for AB and BA patterns. The Hamiltonian of effective *k·p* model can be expressed as:

$$H_k = I_2 \otimes \begin{pmatrix} H^{upper} & H_\perp \\ H_\perp & H^{lower} \end{pmatrix}.$$

Here, $I_2$ is a 2 × 2 identity matrix. $H_\perp$ represents the interlayer hopping term, which is given by:

$$H_\perp = \begin{pmatrix} t_{cc} & 0 \\ 0 & t_{vv} \end{pmatrix},$$

$t_{cc}$ and $t_{vv}$ are the interlayer hopping energies within the conduction and valence bands, respectively. $H^{upper}$ and $H^{lower}$ are the Hamiltonians of the upper and lower layers of the bilayer lattice, respectively, which can be written as:

$$H^{upper(lower)} = H_0^{upper(lower)} + H_{SOC} + (-)[H_{ex} + H_E].$$

Here,

$$H_0^{upper(lower)} = \begin{pmatrix} \frac{\Delta}{2} + \varepsilon & t_{12}(\tau q_x - iq_y) \\ t_{12}(\tau q_x + iq_y) & -\frac{\Delta}{2} + \varepsilon \end{pmatrix},$$

where $\Delta$ represents the band gap at the K (K') valley, $\varepsilon$ is on-site energy, $\tau = \pm 1$ relates to valley index, $\vec{q} = \vec{k} - \vec{K}$ donates the momentum vector of electrons relative to the K (K') valley, and $t_{12}$ is intralayer effective nearest-neighbor hopping integral. The second term, $H_{SOC}$, originating from spin-orbit coupling (SOC) effect can be written as:

$$H_{SOC} = \begin{pmatrix} \tau s \lambda_c & 0 \\ 0 & \tau s \lambda_v \end{pmatrix},$$

where spin index $s = \pm 1$ represents spin up/down, and $\lambda_{c(v)} = E_{c(v)\uparrow} - E_{c(v)\downarrow}$ represents the spin splitting at the bottom of conduction band (the top of the valence band) in single layer due to SOC. The third term, $H_{ex}$, corresponds to the inherent exchange interaction of magnetic ions, which can be described by:

$$H_{ex} = \begin{pmatrix} -sM_c & 0 \\ 0 & -sM_v \end{pmatrix}.$$

Here, $M_{c(v)} = E_{c(v)\downarrow} - E_{c(v)\uparrow}$ represents the effective exchange splitting at the band edge. The fourth term $H_E$ is given by:

$$H_E = I_2 \otimes \frac{U}{2},$$

where $U$ is the electric potential of each single layer.

The low-energy band dispersions around the K and K' valleys for AB and BA patterns obtained from k·p model are illustrated in **Fig. 1(d)**. Obviously, the spontaneous valley polarization is realized in AB and BA patterns. For AB (BA) pattern, the valence band maximum (VBM) at the K (K') valley is from the spin-up (spin-down) channel of upper (lower) layer, while the conduction band minimum (CBM) at the K' (K) valley is from the spin-down (spin-up) channel of the lower (upper) layer. These results are consistent with the former analysis, confirming that the proposed mechanism is accessible. Based on the analysis, it is promising to realize the LP-AHE in the bilayer lattice. For example, when the Fermi level is shifted between the K and K' valleys in the valence band of AB pattern, due to the large Berry curvature at the K valley, the spin-down holes in the upper layer will acquire an transverse velocities ($v \sim E \times \Omega(k)$) and accumulate at the right edge of upper layer in the presence of an in-plane electric field, see **Fig. 1(d)**, to realize the long-sought LP-AHE. In this case, by interlayer-sliding, the AB pattern would switch to the BA pattern, the spin-up holes in lower layer will acquire an transverse velocities and accumulate at the left edge of lower layer due to the opposite Berry curvatures for the K' valleys; see **Fig. 1(d)**. The proposed LP-AHE is therefore ferroelectrically controllable. It should be noted that the LP-AHE can be also obtained when the Fermi level is shifted between K and K' valleys in the conduction band.

One candidate material to validate this mechanism is VSi$_2$P$_4$. **Fig. S1(a)** presents the crystal structure of single-layer VSi$_2$P$_4$, which can be regarded as a 1H-phase VP$_2$ encapsulated by buckled SiP layers. It has a hexagonal lattice with the space group $P\bar{6}m2$. The magnetic ground state of single-layer VSi$_2$P$_4$ is found to be ferromagnetic. A magnetic moment of 1 $\mu_B$ per unit cell is obtained, which is mainly distributed on V atom. **Fig. S1(c)** displays the band structure of single-layer VSi$_2$P$_4$ with SOC in consideration. Clearly, it presents spontaneous valley polarization of 117.6 (22.3) meV in the conduction (valence) band edge. Such features are related to the simultaneous breaking of *T*-symmetry and *P*-symmetry. These results are in consistent with the previous work [29].

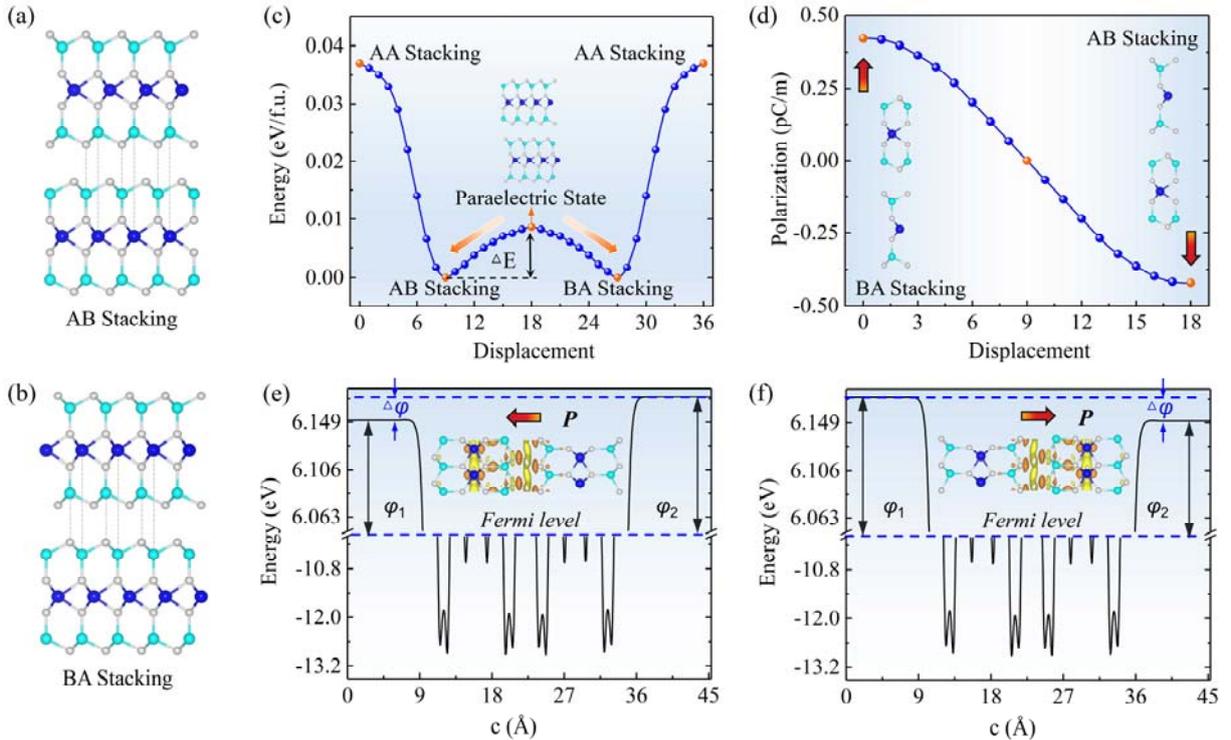

**Fig. 2** Crystal structures of (a) AB- and (b) BA-stacked bilayer VSi$_2$P$_4$. (c) Energy profiles for ferroelectric switching of bilayer VSi$_2$P$_4$. Inset in (c) is the paraelectric state with nonpolar. (d) Variation of out-of-plane electric polarization of bilayer VSi$_2$P$_4$ under ferroelectric switching from BA to AB pattern. Plane averaged electrostatic potentials of (e) AB- and (f) BA-stacked bilayer VSi$_2$P$_4$ along the $z$ direction. Insets in (e) and (f) present the differential charge density diagrams of AB- and BA-stacked bilayer VSi$_2$P$_4$, respectively; yellow and orange isosurfaces represent electron accumulation and depletion, respectively.

Following the proposed mechanism, we construct AB- and BA-stacked bilayer VSi$_2$P$_4$; see **Figs. 2(a)** and **(b)**. Both configurations have the space group of *P*3*m*1, excluding inversion symmetry and $M_z$ symmetry. The optimized lattice constant of bilayer VSi$_2$P$_4$ is 3.46 Å, and the interlayer distance is 3.17 Å. For AB (BA)-stacked bilayer VSi$_2$P$_4$, as shown in **Figs. 2(a)** and **(b)**, the Si (V) atom of upper layer is right above V (Si) atom of lower layer, and the P (Si) atom of upper layer sits directly above Si (P) atom of lower layer, which would result in an out-of-plane electric polarization. This fact is also confirmed by the differential charge density and plane averaged electrostatic potential. As shown in **Figs. 2(e)** and **(f)**, a positive (negative) discontinuity $\Delta V$ = 0.02 (-0.02) eV between the vacuum levels of upper and lower layers clearly indicates the spontaneous out-of-plane electric polarization pointing downward (upward) in AB (BA)-stacked bilayer VSi$_2$P$_4$. And the differential charge density is significantly asymmetric. Based on the Berry phase approach [27], the electric polarization in AB- and BA-stacked bilayer VSi$_2$P$_4$ is calculated to be -0.42 and 0.42 pC/m, respectively.

To determine the ground state for the exchange interaction of bilayer VSi$_2$P$_4$, we consider different magnetic configurations. Our results show that the intralayer and interlayer exchange interactions favor FM and antiferromagnetic (AFM) coupling, respectively, forming the A-type antiferromagnetism. In detail, this state is 0.26 meV/unit cell lower than that with FM interlayer exchange interaction. In AB (BA)-stacked bilayer VSi$_2$P$_4$, due to the existence of electric polarization pointing downward (upward), the magnetic moment distributed on V atom from upper layer is slight larger (smaller) than that from lower layer, which suggests the exotic two-dimensional (2D) magnetoelectric coupling [35-37]. The magnetocrystalline anisotropy energy (MAE), which is defined as the energy difference between the magnetization axis aligned along in-plane and out-of-plane directions, of VSi$_2$P$_4$ bilayer is calculated to be 88.4 ueV/f.u. This indicates that both structures favor out-of-plane magnetization orientation.

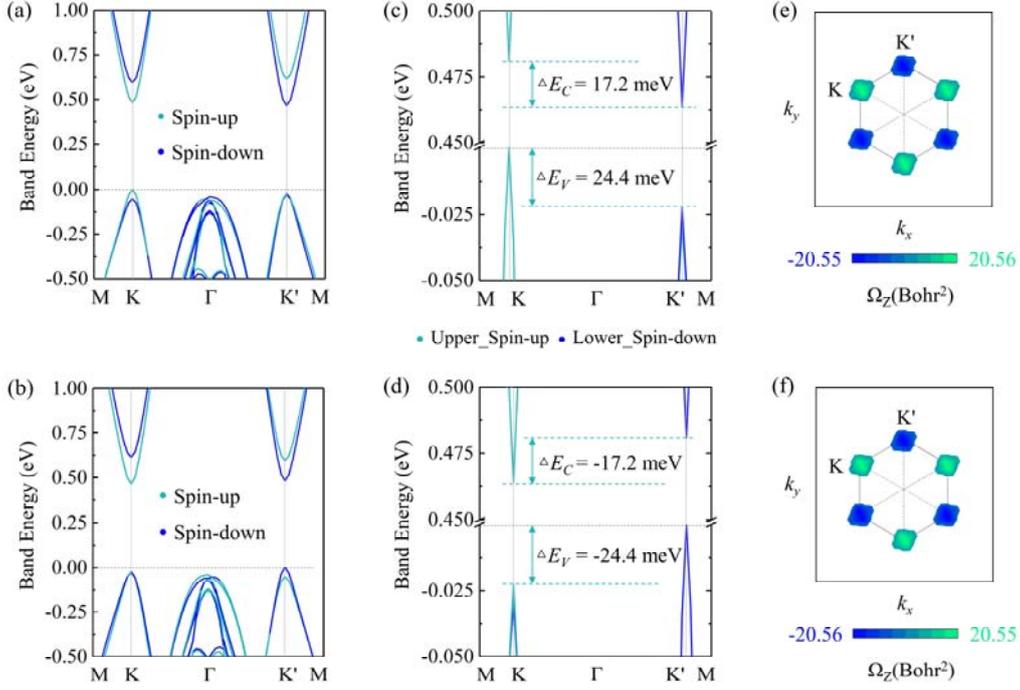

**Fig. 3** Band structures, enlarged low-energy band dispersions and Berry curvatures of (a,c,e) AB- and (b,d,f) BA-stacked bilayer VSi$_2$P$_4$. The Fermi level is set to 0 eV.

**Figs. 3(a)** and **(b)** show the band structures of AB- and BA-stacked bilayer VSi$_2$P$_4$, respectively. AB (BA)-stacked bilayer VSi$_2$P$_4$ exhibits an indirect band gap of 0.46 eV, with the VBM and CBM lying at the K (K') and K' (K) points, respectively. Arising from the existence of out-of-plane electric polarization, for AB (BA) pattern, the VBM (CBM) is from spin-up channel of upper layer, while the CBM (VBM) is from spin-down channel of lower layer; see **Figs. 3(c)** and **(d)**. In this case, the spontaneous valley polarization is realized in both systems, and more importantly, the valleys are layer-locked. The spontaneous valley polarizations in the valence and conduction bands for AB (BA)-stacked bilayer VSi$_2$P$_4$ are calculated to be $\Delta E_V$=24.4meV ($\Delta E_V$=-24.4meV) and $\Delta E_C$=17.2meV ($\Delta E_C$=-17.2meV), respectively. These values are larger than those of the experimentally demonstrated magnetic proximity systems (0.3-1.0 meV)[38,39] and Cr-doped MoSSe (10 meV)[40].

To confirm the layer-locked Berry curvature in AB- and BA-stacked bilayer VSi$_2$P$_4$, we calculate their Berry curvatures using the VASPBERRY code[26]. The Berry curvature is defined as [41]:

$$\Omega(k) = -\sum_n \sum_{n \neq n'} f_n \frac{2 \operatorname{Im} \langle \psi_{nk} | v_x | \psi_{n'k} \rangle \langle \psi_{n'k} | v_y | \psi_{nk} \rangle}{(E_n - E_{n'})^2}.$$

Here, $f_n$ is the Fermi Dirac distribution function, $|\psi_{nk}\rangle$ is the Bloch wave function with eigenvalue $E_n$, and $v_{x(y)}$ is the velocity operator along $x/y$ direction. **Figs. 3(e)** and **(f)** plot the calculated Berry curvatures of AB- and BA-stacked bilayer VSi$_2$P$_4$, respectively. In the following, we take the valence band as an example to discuss the Berry curvature. As shown in **Figs. 3(e)** and **(f)**, for AB (BA) pattern, the Berry curvature at the K (K') valley has a large positive (negative) value. Since the valleys are locked to layers, the layer-locked Berry curvature is obtained. For AB-stacked bilayer VSi$_2$P$_4$, the holes in the K valley would acquire an anomalous velocity $v_a = -\frac{e}{\hbar} E \times \Omega(k)$ [41] in the presence of an in-plane electric field $E$. As shown in **Fig. 4(a)**, by shifting the Fermi level between the K and K' valleys and applying an in-plane electric field, the spin-down holes from the K valley can transversely move to the right edge of upper layer, giving rise to the LP-AHE in AB-stacked bilayer VSi$_2$P$_4$. Different from AB pattern, the holes in the K' valley for BA pattern would gain a reversed anomalous velocity of -$v_a$. When shifting the Fermi level between the K and K' valleys, the spin-up holes from the K' valley would be accumulated at the left edge of the lower layer under an in-plane electric field [**Fig. 4(b)**], forming LP-AHE as well. This scenario is also applicable for the valleys in the conduction band; see **Figs. 4(c)** and **(d)**. Accordingly, the long-sought LP-AHE can be successfully achieved in bilayer VSi$_2$P$_4$.

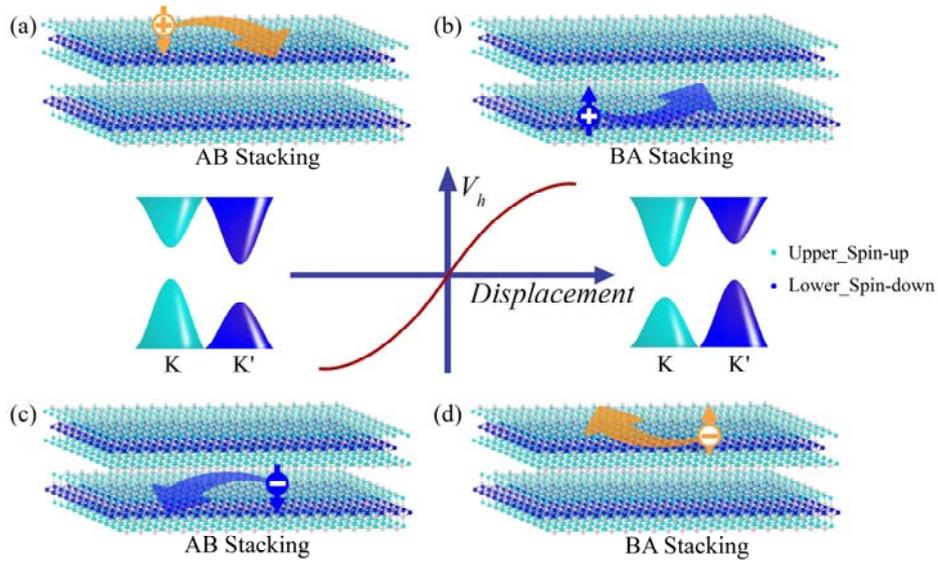

**Fig. 4** Diagrams of the LP-AHE under hole doping for (a) AB- and (b) BA-stacked bilayer VSi$_2$P$_4$. Diagrams of the LP-AHE under electron doping for (c) AB- and (d) BA-stacked bilayer VSi$_2$P$_4$. The holes/electrons from the K and K' valleys are denoted by orange +/- and white +/- symbols, respectively. Arrows refer to the spin directions.

According to the proposed mechanism, the AB- and BA-stacked bilayer VSi$_2$P$_4$ can be considered as two FE states of the sliding ferroelectricity. This indicates that the LP-AHE in bilayer VSi$_2$P$_4$ might be ferroelectrically controllable and reversable, which is highly desirable for device applications. To this end, we investigate the feasibility of its sliding ferroelectricity. The FE switching energy pathway between AB- and BA-stacked bilayer VSi$_2$P$_4$ calculated using the nudge-elastic-band (NEB) method [28] is shown in **Fig. 2(c)**. Due to the rotation symmetry $C_{3z}$, AB pattern can be switched to BA pattern under the interlayer sliding of $t\left[\frac{1}{3},\frac{2}{3},0\right]$, $t\left[-\frac{2}{3},-\frac{1}{3},0\right]$ or $t\left[\frac{1}{3},-\frac{1}{3},0\right]$ (denoted as Path-I). The intermediate state exhibits the space group *Abm*2 (**Fig. S3**) and it is nonpolar due to the existence of glide plane along the *z* direction. From the phonon spectra of the intermediate state shown in **Fig. S4**, imaginary phonon mode can be observed, implying that it is unstable and would experience spontaneous transformation into AB or BA pattern. Besides Path-I, the FE switching from AB to BA patterns can also be achieved under interlayer sliding along the other three equivalent directions, i.e., $t\left[-\frac{2}{3},-\frac{4}{3},0\right]$, $t\left[\frac{4}{3},\frac{2}{3},0\right]$ or $t\left[-\frac{2}{3},\frac{2}{3},0\right]$ (denoted as Path-II). The corresponding intermediate state is the AA pattern. As shown in **Fig. 2(c)**, the FE switching energy barrier along Path-II is found to be 0.037 eV/f.u., while that along Path-I is only 0.009 eV/f.u. Accordingly, the FE switching between AB- and BA-stacked bilayer $VSi_2P_4$ tends to occur along Path-I. The corresponding FE polarizations as a function of step number is shown in **Fig. 2(d)**. The energy barrier is larger than those of T'-$WTe_2$ bilayer ($6 \times 10^{-4}$ eV/f.u.)[42], h-BN bilayer (0.005 eV/f.u.)[36] and β-GeSe bilayer (0.006 eV/f.u.)[43], but lower than that of $In_2Se_3$ (0.066 eV/f.u.)[44]. Such barrier enables the feasibility of FE switching between AB and BA configurations of bilayer $VSi_2P_4$. Therefore, the intriguing FE control of LP-AVH can be realized in bilayer $VSi_2P_4$.

Finally, we wish to emphasize that the proposed mechanism can also be manifested in many other valleytronic van der Waals bilayers, like $VSi_2N_4$, $FeCl_2$, $RuBr_2$, and VClBr, etc. The results are roughly similar to the case of $VSi_2P_4$. The detailed geometric structures, low-energy band dispersions and the Berry curvatures of AB and BA patterns for these systems are shown in **Figs. S5-S9**.

In conclusion, going beyond existing paradigm, we propose a new mechanism to realize the LP-AHE in valleytronic van der Waals bilayers. Through model analysis, we find that the interaction between the out-of-plane ferroelectricity and A-type antiferromagnetism can lead to the layer-locked Berry curvature, enabling the long-sought LP-AHE in the bilayer systems. Moreover, the obtained LP-AHE could couple strongly with sliding ferroelectricity, rendering the LP-AHE ferroelectrically controllable and reversable. Based on first-principles calculations, the validity of this mechanism is demonstrated in a series of real valleytronic materials, including bilayer $VSi_2P_4$, $VSi_2N_4$, $FeCl_2$, $RuBr_2$ and VClBr.

# AUTHOR INFORMATION

Corresponding Authors

*E-mail: daiy60@sina.com (Y.D.).

*E-mail: yandong.ma@sdu.edu.cn (Y.M.).

ORCID

Ying Dai: 0000-0002-8587-6874

Yandong Ma: 0000-0003-1572-7766

Notes

The authors declare no competing financial interest.

# SUPPORTING INFORMATION

Supporting Information Available: crystal structures and band dispersions of single-layer $VSi_2P_4$, AA-stacked bilayer $VSi_2P_4$ and paraelectric state, phonon spectra of the paraelectric state, and detailed results for bilayer $VSi_2P_4$, $VSi_2N_4$, $FeCl_2$, $RuBr_2$ and VClBr.

# ACKNOWLEDGMENTS

This work is supported by the National Natural Science Foundation of China (No. 11804190, 12074217), Shandong Provincial Natural Science Foundation (Nos. ZR2019QA011 and ZR2019MEM013), Shandong Provincial Key Research and Development Program (Major Scientific and Technological Innovation Project) (No. 2019JZZY010302), Shandong Provincial Key Research and Development Program (No. 2019RKE27004), Shandong Provincial Science Foundation for Excellent Young Scholars (No. ZR2020YQ04), Qilu Young Scholar Program of Shandong University, and Taishan Scholar Program of Shandong Province.